\documentstyle[aps,epsf]{revtex}           
\newcommand{\postscript}[2] {\setlength{\epsfxsize}{#2\hsize} 
\centerline{\epsfbox{#1}}} 
\begin{document} 
\title{Virtual states of light non-Borromean halo nuclei} 
\author{ A.Delfino$^{1}$, T. Frederico$^{2}$, M.S.Hussein$^{3}$
Lauro Tomio$^{4}$} 
\address{ 
$1$Departamento de F\'\i sica, Universidade Federal Fluminense,
\\ 24210-340  Niter\'oi, Rio de Janeiro, Brasil.\\
$^{2}$ Dep. de F\'\i sica, Instituto Tecnol\'ogico de Aeron\'autica,
Centro T\'ecnico Aeroespacial,\\ 
12228-900 S\~ao Jos\'e dos Campos, Brasil \\
$^{3}$ Nuclear Theory and Elementary Particle Phenomenology
Group,\\ Instituto de F\'{\i}sica, Universidade de S\~{a}o Paulo,
05315-970, S\~{a}o Paulo, Brasil \\
$^{4}$ Instituto de F\'\i sica Te\'orica, Universidade Estadual
Paulista, 01405-900 S\~{a}o Paulo, Brasil \\ } 
\date{\today} 
\maketitle 
\begin{abstract} 
It is shown that the three-body non-Borromean halo nuclei like $^{12}Be$,
$^{18}C$, $^{20}C$ have $p-$wave virtual states with energy of about 1.6
times the corresponding neutron-core binding energy. We use a
renormalizable model that guarantees the general validity  of our results in
the context of short range interactions. 
\end{abstract} 
\vskip 1cm
 
Halo nuclei offer the opportunity to study the few-body
aspects of the nuclear interaction with their peculiar
three-body phenomena. Recently, attention was drawn to the possibility of
existing Efimov states\cite{ef70} in such systems, because some halo
nuclei can be viewed as a three body system with two loosely bound
neutrons and a core~\cite{tani,hu93,zhu93}. It was suggested~\cite{fe93} 
$^{18}C$ and $^{20}C$ as promising candidates to have Efimov states. 
In Ref.~\cite{am97}, by considering the critical conditions to allow the
existence of one Efimov state, using the experimental values for the
neutron separation energies ($^{19}C+n$ and  $^{18}C+2n$) 
given in \cite{audi}, it was concluded that $^{20}C$ could have such state. 

The weakly bound  Efimov states\cite{ef70} appear in the
zero angular momentum state of a  three boson system and the number of
states grows to infinity, condensing at zero energy as the pair interactions 
are just  about to bind two particles in $s-$wave. Such  states are
loosely 
bound and their  wave functions extend far beyond those of normal states.
If such states exist in nature they  
will dominate the low-energy scattering of one of the particles with the 
bound-state of the remaining two particles. Such states have been studied in 
several numerical model calculations\cite{fe93,efim2,tom}.  
There were theoretical searches  for Efimov states in atomic and nuclear 
systems without a clear experimental signature of their
occurrence.~\cite{evid1,ef90,am99}.  

The physical picture underlying such phenomena is related to the unusually large
size of these light three-body halo nuclei. The core can be assumed structureless 
\cite{fe93,das94}, considering that the radius of the neutron halo is much
greater than the radius of the core. \ The large size scale of the orbit of the
outer neutrons in halo nuclei comes from the small neutron separation energies,
characterizing a weakly bound few-body system. Thus, the detailed form of the
nuclear interaction is not important giving to the system universal properties as
long some physical scales are known \cite{am97}. This situation allows the use of
concepts coming from short-range interactions. 

In the limit of a zero-range interaction the three-body system is parameterized by
the physical two-body and three-body scales. In a renormalization approach of the
quantum mechanical many-body model with the $s-$wave zero-range force,
all the 
low-energy properties of the three-body system are well defined if one three-body
and another two-body physical informations are known~\cite{ad95}. The
three-body input can be chosen as the experimental ground state binding energy.
All the detailed informations about the short-range force, beyond the low-energy
two-body observables, are retained in only one three-body physical information in
the limit of  zero range interaction.  The sensibility of the three body binding
energy to the interaction properties comes from the collapse of the system in the  
limit of zero-range force, which is known as the Thomas effect~\cite{th35}.  

The three-body scale vanishes as a physical parameter if angular momentum or
symmetry do not allow the simultaneous presence of the particles close to each
other. In three-body $p-$wave states, the particles interacting 
through $s-$wave potentials have the centrifugal barrier forbbiden the third 
particle to  be close to the interacting pair. 
Consequently, the third particle just notice the
asymptotic wave of the interacting pair, which is defined  by a two-body physical
scale, and the three-body scale is not seen by the system in these states. The 
observables of the three-body system in states which have non zero angular
momentum are determined just by two-body scales. We look for special possibilities
in $p-$wave like the virtual state. The trineutron system in $p-$wave presents a
peculiar pole in the second energy sheet~\cite{Glockle,delf96},  when the
neutron-neutron ($n-n$) is artificially bound. The value of the pole scale with
the binding energy of the fictious $n-n$ system, as this is the only scale
of the
three-body system~\cite{delf96}. It is not forbbiden, in principle, to exist 
one virtual state of the three-body halo nuclei system in $p-$wave, and
if it exists, it depends exclusively on the two-body scales: the binding 
energy of the neutron to the core and the $n-n$ virtual state energy.
 
In this work, we search for the virtual state of the three-body halo nuclei
in $p-$wave. We make use of the zero-range model which is well defined
in $p-$wave, and the inputs are the energy of the bound state of the
neutron to the core and the virtual $n-n$ state energy. We look for
weakly bound $n-$core systems, in particular we look at 
$^{12}Be$ ($^{10}Be+2n$), $^{18}C$ ($^{16}C+2n$), and $^{20}C$ ($^{18}C+2n$).
The zero-range model is analitically continued to the second sheet, in the 
complex energy plane, and there we seek for the solution of the homogeneous
equation.  In the case of Borromean  halo nuclei such as $^{11}Li$, our
method does not work.  However to get reed of the virtual state, the
$^{10}Li$ core is made artificially bound, to allow the analytical
continuation to the second energy sheet through the elastic cut.

The nuclei $^{12}Be$, $^{18}C$, and $^{20}C$ have an interesting non-Borromean
nature with strong $n-n$ pairing in the ground-state. Specifically, 
$^{12}Be$ is $\left\{ O^+, 23.6 {\rm ms}, E_n=3169 {\rm KeV} \right\}$,
$^{18}C$ is $\left\{ O^+, 95 {\rm ms}, E_n=4180 {\rm KeV} \right\}$, and
$^{20}C$ is $\left\{ O^+, \;\;?\;\; , E_n=3340 {\rm KeV} \right\}$, where
the first number is the spin-parity of the ground state, the second is the
mean lifetime and the third is the neutron separation energy. 
The lifetime of $^{20}C$, shown by a question mark, is not available. 
The numbers should be
compared to the one-neutron-less isotopes, $^{11}Be$, $^{17}C$, and $^{19}C$,
respectively given by $\left\{ 1/2^+, 13.81 {\rm ms}, E_n=504 {\rm KeV}
\right\}$, $\left\{ \;\;?\;\;, 193 {\rm ms}, E_n=729 {\rm KeV} \right\}$,
and $\left\{ 5/2^+  (1/2^+), \;\;?\;\;, E_n=160 (530) {\rm KeV} \right\}$.
The number in the round brackets, in $^{19}C$ refer to the recent
measurement of Nakamura et al.~\cite{naka}. Again, the question marks
refer to not available results. 
The above nuclei are used to determine the neutron-core
binding energies in our calculation to follow.  \ Note that the $n-n$ pairing
energies $\Delta_{nn}$ are in the range $2260\le\Delta_{nn}\le 3400$ KeV. \ 
In our calculation of the $p-$wave virtual state, the pairing is taken
inoperative and the only energy scales left are the neutron-core binding 
energy $(E_{nc})$ and the $n-n$ virtual state energy $(E_{nn})$ in the
$p-$wave three-body virtual state (pygmy dipole state).

As the input energies  are fixed in the renormalized model, a more
realistic potential will not affect the generality of the present
conclusions. The Pauli principle correction, between the halo and  the
core neutrons, affects essentially the ground state and it is weakened in
the $p-$wave state due to the centrifugal barrier. We have to consider
that
this is a short-range phenomenon that occurs for distances less than the
core size (about $\approx 3 fm$ for light-halo nuclei).  We believe that
our results are valid  even in the case where the spin of the core is non
zero. The results show little dependence on the mass difference
of the particles, in a sense explained together with the numerical
results, enough to indicate that the dependence on the details of the
interactions cannot be larger.  
 
In other context, the three-nucleon system has been studied with  
zero-range force models \cite{ef90}.  
They succeeded in explaining  
the qualitative properties of the  
three-nucleon system and described the known
correlations between three-nucleon observables. 
The universality in the 
three-nucleon system  means the independence of the correlations 
to the details of the short-range nucleon-nucleon potentials \cite{ef90}. 
 
Here we use a notation appropriate for halo nuclei, $n$ for neutron 
and $c$ for core, but we would like to point out that our approach  
is applicable  to any three-particle system that interact  
via $s-$wave short-range interactions,  where two of the particles  
are identical.  The $s-$wave interaction for the $n-c$ potential is  
justified in the present analysis, because the $p-$wave virtual state if
exists  
it should have a small energy, just being sensitive to the 
properties of the zero angular  
momentum two-particle  state in the relative coordinates.
It also was observed in Ref.\cite{zin}, when discussing $^{11}Li$, that  
even the three-body wave-function with an $s-$wave $n-n$ correlation  
produces a ground  state of the halo nuclei with 
two or more shell-model configurations.   
 
The energies of the two particle subsystem, $E_{nn}$ and $E_{nc}$ can be  
virtual or bound. However, the extension to the second energy sheet, will
be done through the cut of the elastic scattering of the neutron and the
bound neutron-core subsystem. Thus, we are going to use the value of
the virtual state energy $E_{nn}$=143 KeV and the binding energy of the
neutron to the core $E_{nc}$ in our calculations.
We vary the core mass to study the light halo nuclei  
 like $^{11}Li$, $^{12}Be$, $^{18}C$ and $^{20}C$. 
 
The zero-range three-body integral equations for the bound state of two 
identical particles and a core, is written as  
generalization of the three-boson equation \cite{st57}. It is 
composed by two coupled integral equations in close analogy to the case of 
$s-$wave separable potential model presented in Ref.\cite{das94}.  
The antisymmetrization of the two outer neutrons  
is satisfied since the spin couples to zero \cite{fe93}. 
In our approach the potential form factors and 
corresponding strengths are replaced, in the 
renormalization procedure, by the two-body binding energies,  
$E_{nn}$ and $E_{nc}$.  
In the case of bound systems, these quantities are the separation energies. 
We distinguish these two cases  by the following definition:   
\begin{equation} 
K_{nn} \equiv  -\sqrt{E_{nn}} , \; \; \;  
K_{nc} \equiv \sqrt{E_{nc}} ,  
\label{K} 
\end{equation}  
where $+$ refers to bound and $-$ to virtual state-energies. 
Our units will be such that $\hbar = 1$ and the nucleon mass, $m_{n} = 1$.  
 
After partial wave projection, the $\ell-$wave coupled integral equations  
for the three-body system consisting of two neutrons and a core ($n-n-c$)
are: 
\begin{eqnarray} 
\chi^\ell_{nn}(q)&=&2\tau_{nn}(q;E;K_{nn}) 
\int_0^\infty dk G_1^\ell(q,k;E)\chi^\ell_{nc}(k)  
\label{chi1} \\  
\chi^\ell_{nc}(q)&=&\tau_{nc}(q;E;K_{nc})\int_0^\infty dk  
\left[G^\ell_1(k,q;E) \chi^\ell_{nn}(k)  
+  A_c G^\ell_2(q,k;E) \chi^\ell_{nc}(k)\right] , 
\label{chi2}  
\end{eqnarray} 
where 
\begin{eqnarray} 
\tau_{nn}(q;E;K_{nn})&=& 
\frac{1}{\pi} 
\left[\sqrt{E+\frac{A_c+2}{4A_c} q^2}-K_{nn}\right]^{-1}, 
\label{tau1} \\  
\tau_{nc}(q;E;K_{nc})&=&\frac{1}{\pi}\left(\frac{A_c+1}{2A_c}\right)^{3/2} 
\left[\sqrt{E+\frac{A_c+2}{2(A_c+1)} q^2}-K_{nc}\right]^{-1}, 
\label{tau2} 
\end{eqnarray} 
\begin{eqnarray} 
G^\ell_1(q,k;E)&=&2A_c k^2
\int^1_{-1}dx \frac{P_\ell(x)}
{2A_c(E+k^2)+q^2(A_c+1)+2A_c qkx}, 
\label{G1} \\  
G^\ell_2(q,k;E)&=&2 k^2
\int^1_{-1}dx \frac{P_\ell(x)}
{2A_cE+(q^2+k^2)(A_c+1)+2qkx} .
\label{G2}  
\end{eqnarray} 
 
In the above equations, $A_c$ is the core mass number and $E$ is modulus of
the energy of the three-body halo state.  As we are interested in the $\ell$-th
angular momentum three-body state, the Thomas collapse is forbidden
and the integration over momentum extends to infinity. For $\ell > 0$
the short-range three-body scale are not seen by the system, while  for
 $\ell = 0$ the
renormalization of the Faddeev equations is necessary. In the renormalization
procedure, for
$\ell=0$ a subtraction should be performed in the Faddeev equations 
and the momentum scale, which
represents the subtraction point in the integral equation \cite{ad95},
qualitatively  represents the inverse of the interaction 
radius \cite{ef90}. The subtraction point goes 
to infinity as the radius of the interaction decreases.  
The three-body model is renormalizable for $\ell=0$,   
requiring only one three-body observable to be fixed\cite{ad95}, 
which is the physical meaning of the subtraction performed in the
Faddeev equations,
together with the two-body low-energy physical informations. 
The  scheme is invariant under renormalization group transformations.
However, for $\ell >0$, the original equations as given by (\ref{chi1})
and (\ref{chi2}) are well defined and the three-body observables
are completly determined by the two-body physical scales corresponding to
$K_{nn}$ and $K_{nc}$.
 
The analytic continuation of the scattering equations for
separable potentials to the second energy sheet, has been
extensively discussed by Gl\"ockle\cite{Glockle} and, in the case of the
zero-range three-body model \cite{st57}, by Frederico et al.\cite{virtual}.
The integral equations on the second energy sheet are obtained by 
the analytical continuation  through the two-body elastic scattering cut 
due to neutron scattering on the bound  neutron-core  subsystem. 
The elastic scattering cut comes through the pole of the neutron-core 
elastic scattering amplitude in Eq.(\ref{tau2}). 
In the next, we perform the analytic continuation of Eqs.
(\ref{chi1} - \ref{G2}) to the second energy sheet. 
The spectator function $\chi^\ell_{nc}(k)$ is substituted by 
$\chi^\ell_{nc}(k) / \left[E_v-E_{nc}+
\displaystyle{\frac{A_c+2}{2(A_c+1)}}k^2\right]$, where
$E_v$ is the modulus of the virtual state energy.
The resulting coupled equations in the second energy sheet are given by:

\begin{eqnarray} 
\chi^\ell_{nn}(q)&=&\tau_{nn}(q;E_v;K_{nn}) 
 \frac{4i(A_c+1)}{\pi q(A_c+2)}
G_1^\ell(q,-ik_v;E_v)\chi^\ell_{nc}(-ik_v)  \nonumber \\
&+&2\tau_{nn}(q;E_v;K_{nn}) 
\int_0^\infty dk \frac{G_1^\ell(q,k;E_v)\chi^\ell_{nc}(k)}
{E_v-E_{nc}+{\displaystyle{\frac{A_c+2}{2(A_c+1)}}}k^2} \ ,
\label{vchi1} \\  
\chi^\ell_{nc}(q)&=&\overline{\tau}_{nc}(q;E_v;K_{nc})
 \frac{2iA_c(A_c+1)}{\pi q(A_c+2)}
G_2^\ell(q,-ik_v;E_v)\chi^\ell_{nc}(-ik_v)
\nonumber \\
&+&\overline{\tau}_{nc}(q;E_v;K_{nc})
\int_0^\infty dk  
\left(G^\ell_1(k,q;E_v) \chi^\ell_{nn}(k)  
+ \frac{A_c G^\ell_2(q,k;E_v) \chi^\ell_{nc}(k)}{E_v-E_{nc}+
{\displaystyle{\frac{A_c+2}{2(A_c+1)}}}k^2
}\right) , 
\label{vchi2}  
\end{eqnarray} 
where, the on-energy-shell momentum at the virtual state is
$k_v=\sqrt{\displaystyle{\frac{2(A_c+1)}{A_c+2}}(E_v-E_{nc})}$, and
\begin{eqnarray} 
\overline{\tau}_{nc}(q;E;K_{nc})&=&
\frac{1}{\pi}\left(\frac{A_c+1}{2A_c}\right)^{3/2} 
\left[\sqrt{E+\frac{A_c+2}{2(A_c+1)} q^2}+K_{nc}\right], 
\label{vtau2} 
\end{eqnarray} 

The cut of the elastic amplitude, given by the exchange of the
core between the different possibilities of the bound core-neutron 
subsystems, is near the physical region of virtual state pole, 
due to the small value of $E_{nc}$. This cut is given by the values 
of imaginary $k$ between the extreme poles of the free 
three-body Green's function, $G^\ell_2(q,k;E_v)$, given by
Eq.(\ref{G2}), which appears in the first term of the right-hand side of
Eq.(\ref{vchi2}),
\begin{eqnarray}
2A_cE+(q^2+k^2)(A_c+1)+2qkx = 0 
,\label{cut}
\end{eqnarray} 
with $-1<x<1$, $q=k=-i k_{cut}$ and
 $E=\displaystyle{\frac{A_c+2}{2(A_c+1)}} k_{cut}^2+E_{nc}$. Introducing
the value of $E$ substituting the imaginary $k$ in Eq.(\ref{cut}), the cut
is found at values of $E$ satisfying
\begin{eqnarray}
2\frac{A_c+1}{A_c}E_{nc} \ > \ E \ > \ 2\frac{A_c+1}{A_c+2}E_{nc} \ .
\label{cut1}
\end{eqnarray} 
The virtual state energy $E_v$ in the second energy sheet is
found  between the scattering threshold and the cut,
$E_v <  2\displaystyle{\frac{A_c+1}{A_c+2}}E_{nc}$, which gives 
for $B_v =E_v-E_{nc}<\displaystyle{\frac{A_c}{A_c+2}}E_{nc}$. 

In the limit of zero-ranged interaction the only physical 
scales of the three-body system for $p-$waves is $E_{nn}$ and $E_{nc}$, 
implying that $B_v= E_{nc}{\cal F}\left(
\displaystyle{{E_{nc}}/{E_{nn}}},A_c\right)$,
where $\cal{F}$ is a scaling function to be determined by the
solution of Eqs. (\ref{vchi1}) and (\ref{vchi2}).
However, because of the proximity of the cut to the scattering
threshold, it is reasonable to believe that it should have
a major importance for the formation of the virtual state, and 
${\cal F}({E_{nc}}/{E_{nn}},A_c)$ be roughly independent of the
ratio $E_{nc}/E_{nn}$. Another consequence of the
dominance of the cut in the virtual state energy, should be
a soft dependence on $A_c$ of the ratio  $B_v(A_c+2)/(E_{nc}A_c)$. 

In figure 1, the results of the virtual state energy are
shown in the form of the ratio $B_v(A_c+2)/(E_{nc}A_c)$ as a function
of the core mass $A_c$, for $E_{nc}=E_{nn}$. The numerical values  
of the virtual $n-n$ and bound $n-c$ states energies can be choosen as
being equal. The calculation are shown for a extreme variation of
$A_c$ between .001 and 1000, while the ratio changed by a factor of
three. The other characteristic  of the virtual state 
is the approximate independence of $B_v(A_c+2)/(E_{nc}A_c)$ 
on the ratio $E_{nc}/E_{nn}$, which is confirmed in figure 2, where
calculations were performed for $E_{nc}/E_{nn}$ between .01 and 1000.

The three-body halo nuclei $^{11}Li$, $^{12}Be$, $^{18}C$ and
$^{20}C$ have the $p-$wave virtual state. In the case of $^{11}Li$
we artificially changed the virtual state of $^{10}Li$ to a bound state,
just to give to the reader one value of the three-body virtual state
in case the binding energy of the neutron to the core is some tenth's of 
KeV. In table I, our results are show. The $p-$wave virtual  state energy
scales with the binding energy of the neutron to the core, and it
is roughly of about twice $E_{nc}$. 

In summary, we have discussed the universal aspects of $p-$wave virtual
states of three-body halo nuclei 
in the limit of a zero-range interaction.  We show the existence of 
scaling properties of the three-body $p-$wave virtual state energy in
respect  to the energies of the  $n-n$ virtual  and $n-c$ bound states
which  
determine the value of the $p-$wave virtual state energy. We conclude that
the scaling function ${\cal F}({E_{nc}}/{E_{nn}},A_c)$ which gives
the virtual state energy as 
$E_v=E_{nc}\left[1+{\cal F}({E_{nc}}/{E_{nn}},A_c)\right]$, is rougly
independent
on the ratio $E_{nc}/E_{nn}$, and  approximately entirely determined by
$A_c$. \ From knowledge of $E_{nc}$, we calculated the $p-$wave virtual state
energies for $^{12}Be$, $^{18}C$ and $^{20}C$ which came out to be about 1.6
times the neutron-core binding energy.
These threshold dominated excited states, commonly called ``pygmy
resonances", are therefore not resonances at all and correspond to a
manifestation of predominantly dipole final state interaction, just as in
the two-body case of the most well known halo nucleus, the
deuteron~\cite{HLT}. 

\begin{table} 
\caption[dummy0]
{$p-$wave virtual state energy of light-halo nuclei. $E_{nc}$ is the
binding  energy of the neutron to the core, considering the central value
given in Ref.~\cite{audi}. For $^{20}C $ we also use another value for the
binding energy of a neutron to $^{19}C $ ($E_{nc}=530\pm 130$ KeV), 
as given in Ref~\cite{naka}.}
\begin{center} 
\begin{tabular}{ccccc} 
$Nucleus$&$A_c$& $E_{nc}$ (KeV) &  $E_v$ (KeV) &$B_v(A_c+2)/(E_{nc}A_c)$\\ 
\hline  
$^{11}Li$&9  & 50\  &  79.54\   & 0.7221\\ 
$^{12}Be$&10 &504\  & 813.65\   & 0.7373\\ 
$^{18}C $&16 &729\  &1227.54\   & 0.7694\\ 
$^{20}C $&18 &162\  & 274.36\   & 0.7706\\ 
$^{20}C $&18 &530\  & 900.30\   & 0.7763\\ 
\end{tabular} 
\end{center} 
\end{table} 

Our thanks for support from Funda\c c\~ao de Amparo \`a Pesquisa do Estado  
de S\~ao Paulo (FAPESP) and from Conselho Nacional de  Desenvolvimento
Cient\'{\i}fico e Tecnol\'ogico (CNPq) of Brazil.

\begin{figure}
\postscript{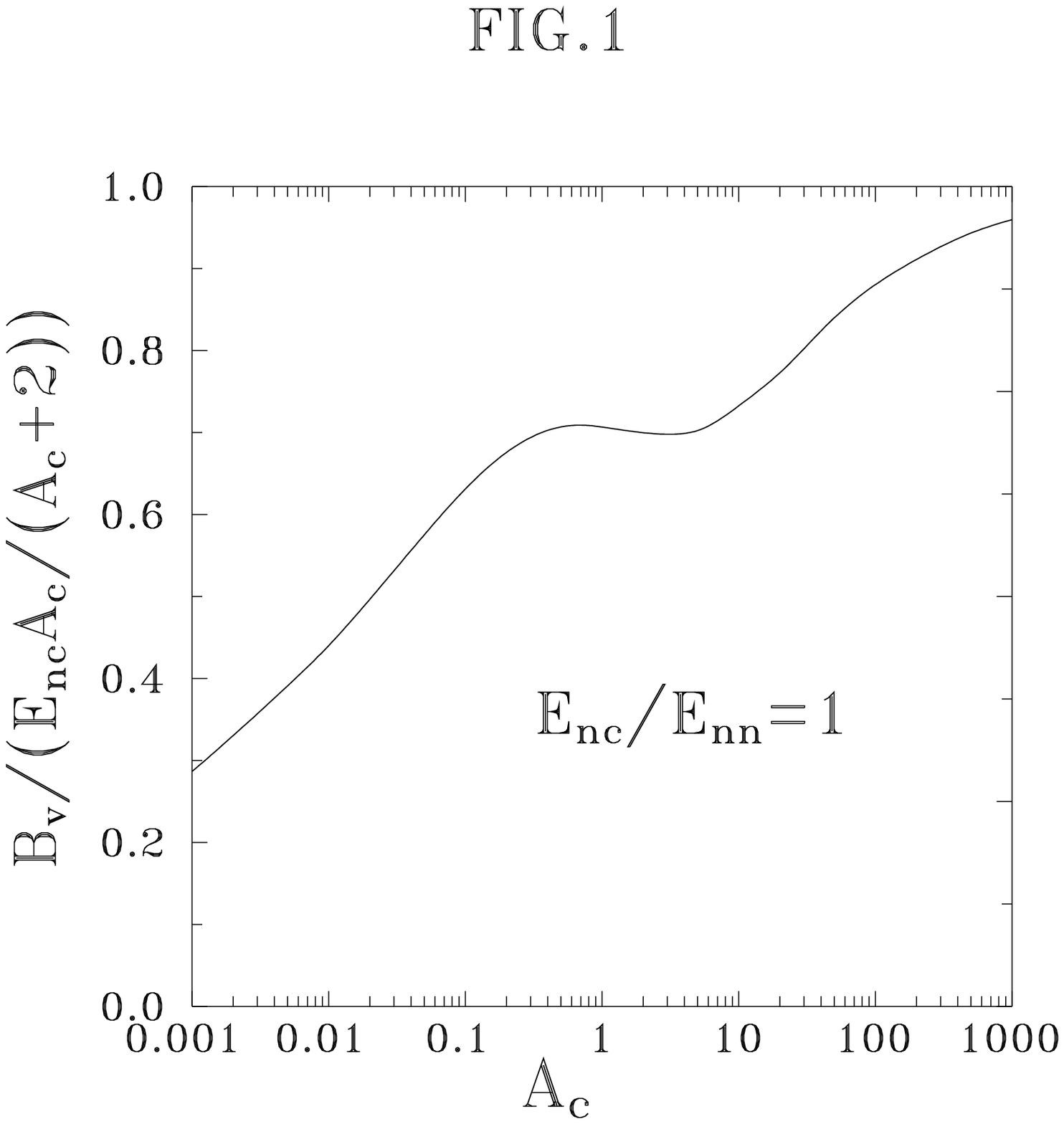}{0.8}    
\caption{Scaling plot of 
$\displaystyle{\frac{A_c+2}{A_c}\frac{B_v}{E_{nc}}}$ 
as a function of  $A_c$ for $E_{nc}=E_{nn}$.}
\end{figure}
\begin{figure}
\vspace{1 true cm} 
\postscript{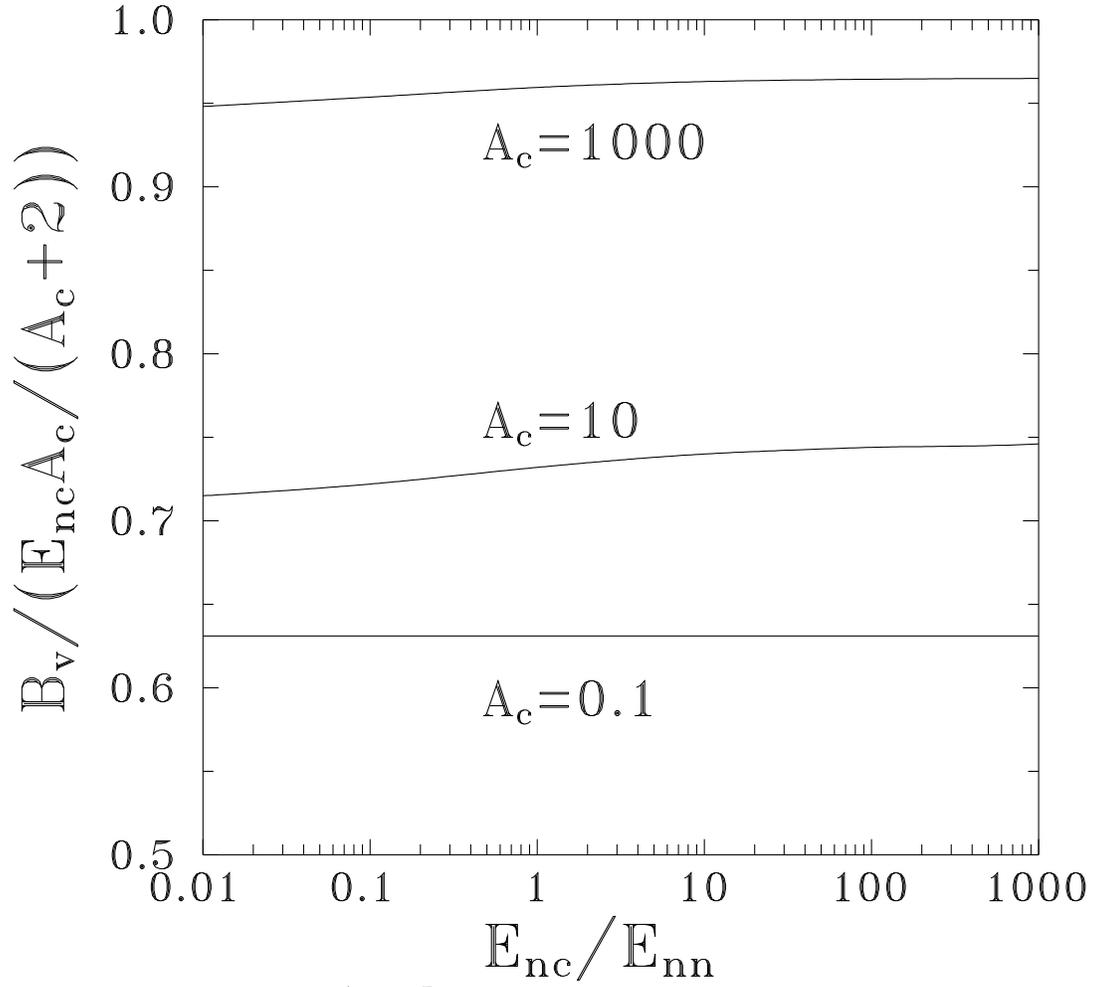}{0.8}    
\caption{Scaling plot of 
$\displaystyle{\frac{A_c+2}{A_c}\frac{B_v}{E_{nc}}}$ 
as a function of $E_{nc}/E_{nn}$ for $A_c=$ 0.1 , 10 and 100.} 
\end{figure}
\newpage
\centerline{Figure Captions}
\vskip 1cm
\begin{description}
\item[Fig. 1]  
Scaling plot of 
$\displaystyle{\frac{A_c+2}{A_c}\frac{B_v}{E_{nc}}}$ 
as a function of  $A_c$ for $E_{nc}=E_{nn}$.
\vskip 0.5cm
\item[Fig. 2]  
Scaling plot of 
$\displaystyle{\frac{A_c+2}{A_c}\frac{B_v}{E_{nc}}}$ 
as a function of $E_{nc}/E_{nn}$ for $A_c=$ 0.1 , 10 and 100.
\end{description}
\end{document}